\documentstyle[prd,aps,floats,preprint]{revtex}      

\input psfig                           
\begin{document}
\draft
\tighten                    

\preprint{\vbox{\hbox{\bf CLNS 96/1435 \hfill}
                \hbox{\bf CLEO 96-18   \hfill}
                \hbox{\bf \today       \hfill}}}

\title{\bf Measurement of the Direct Photon Spectrum in
           $\Upsilon({\rm 1S})$ Decays}
\author{
B.~Nemati, S.~J.~Richichi, W.~R.~Ross, P.~Skubic, and M.~Wood}
\address{
University of Oklahoma, Norman, Oklahoma 73019}
\author{
M.~Bishai, J.~Fast, E.~Gerndt, J.~W.~Hinson, N.~Menon,
D.~H.~Miller, E.~I.~Shibata, I.~P.~J.~Shipsey, and M.~Yurko}
\address{
Purdue University, West Lafayette, Indiana 47907}
\author{
L.~Gibbons, S.~D.~Johnson, Y.~Kwon, S.~Roberts, and E.~H.~Thorndike}
\address{
University of Rochester, Rochester, New York 14627}
\author{
C.~P.~Jessop, K.~Lingel, H.~Marsiske, M.~L.~Perl,
S.~F.~Schaffner, D.~Ugolini, R.~Wang, and X.~Zhou}
\address{
Stanford Linear Accelerator Center, Stanford University, Stanford,
California 94309}
\author{
T.~E.~Coan, V.~Fadeyev, I.~Korolkov, Y.~Maravin, I.~Narsky,
V.~Shelkov, J.~Staeck, R.~Stroynowski, I.~Volobouev, and J.~Ye}
\address{
Southern Methodist University, Dallas, Texas 75275}
\author{
M.~Artuso, A.~Efimov, F.~Frasconi, M.~Gao, M.~Goldberg, D.~He,
S.~Kopp, G.~C.~Moneti, R.~Mountain, Y.~Mukhin, S.~Schuh,
T.~Skwarnicki, S.~Stone, G.~Viehhauser, and X.~Xing}
\address{
Syracuse University, Syracuse, New York 13244}
\author{
J.~Bartelt, S.~E.~Csorna, V.~Jain, and S.~Marka}
\address{
Vanderbilt University, Nashville, Tennessee 37235}
\author{
A.~Freyberger, D.~Gibaut, R.~Godang, K.~Kinoshita, I.~C.~Lai,
P.~Pomianowski, and S.~Schrenk}
\address{
Virginia Polytechnic Institute and State University,
Blacksburg, Virginia 24061}
\author{
G.~Bonvicini, D.~Cinabro, R.~Greene, and L.~P.~Perera}
\address{
Wayne State University, Detroit, Michigan 48202}
\author{
B.~Barish, M.~Chadha, S.~Chan, G.~Eigen, J.~S.~Miller,
C.~O'Grady, M.~Schmidtler, J.~Urheim, A.~J.~Weinstein,
and F.~W\"{u}rthwein}
\address{
California Institute of Technology, Pasadena, California 91125}
\author{
D.~M.~Asner, D.~W.~Bliss, W.~S.~Brower, G.~Masek, H.~P.~Paar,
M.~Sivertz, and  V.~Sharma}
\address{
University of California, San Diego, La Jolla, California 92093}
\author{
J.~Gronberg, R.~Kutschke, D.~J.~Lange, S.~Menary, R.~J.~Morrison,
H.~N.~Nelson, T.~K.~Nelson, C.~Qiao, J.~D.~Richman, D.~Roberts,
A.~Ryd, and M.~S.~Witherell}
\address{
University of California, Santa Barbara, California 93106}
\author{
R.~Balest, B.~H.~Behrens, K.~Cho, W.~T.~Ford, H.~Park, P.~Rankin,
J.~Roy, and J.~G.~Smith}
\address{
University of Colorado, Boulder, Colorado 80309-0390}
\author{
J.~P.~Alexander, C.~Bebek, B.~E.~Berger, K.~Berkelman, K.~Bloom,
D.~G.~Cassel, H.~A.~Cho, D.~M.~Coffman, D.~S.~Crowcroft,
M.~Dickson, P.~S.~Drell, K.~M.~Ecklund, R.~Ehrlich, R.~Elia,
A.~D.~Foland, P.~Gaidarev, R.~S.~Galik,  B.~Gittelman,
S.~W.~Gray, D.~L.~Hartill, B.~K.~Heltsley, P.~I.~Hopman,
J.~Kandaswamy, N.~Katayama, P.~C.~Kim, D.~L.~Kreinick, T.~Lee,
Y.~Liu, G.~S.~Ludwig, J.~Masui, J.~Mevissen, N.~B.~Mistry,
C.~R.~Ng, E.~Nordberg, M.~Ogg,%
\thanks{Permanent address: University of Texas, Austin TX 78712}
J.~R.~Patterson, D.~Peterson, D.~Riley, A.~Soffer, and C.~Ward}
\address{
Cornell University, Ithaca, New York 14853}
\author{
M.~Athanas, P.~Avery, C.~D.~Jones, M.~Lohner, C.~Prescott,
S.~Yang, J.~Yelton, and J.~Zheng}
\address{
University of Florida, Gainesville, Florida 32611}
\author{
G.~Brandenburg, R.~A.~Briere, Y.S.~Gao, D.~Y.-J.~Kim, R.~Wilson,
and H.~Yamamoto}
\address{
Harvard University, Cambridge, Massachusetts 02138}
\author{
T.~E.~Browder, F.~Li, Y.~Li, and J.~L.~Rodriguez}
\address{
University of Hawaii at Manoa, Honolulu, Hawaii 96822}
\author{
T.~Bergfeld, B.~I.~Eisenstein, J.~Ernst, G.~E.~Gladding,
G.~D.~Gollin, R.~M.~Hans, E.~Johnson, I.~Karliner, M.~A.~Marsh,
M.~Palmer, M.~Selen, and J.~J.~Thaler}
\address{
University of Illinois, Champaign-Urbana, Illinois 61801}
\author{
K.~W.~Edwards}
\address{
Carleton University, Ottawa, Ontario, Canada K1S 5B6 \\
and the Institute of Particle Physics, Canada}
\author{
A.~Bellerive, R.~Janicek, D.~B.~MacFarlane, K.~W.~McLean,
and P.~M.~Patel}
\address{
McGill University, Montr\'eal, Qu\'ebec, Canada H3A 2T8 \\
and the Institute of Particle Physics, Canada}
\author{
A.~J.~Sadoff}
\address{
Ithaca College, Ithaca, New York 14850}
\author{
R.~Ammar, P.~Baringer, A.~Bean, D.~Besson, D.~Coppage,
C.~Darling, R.~Davis, N.~Hancock, S.~Kotov, I.~Kravchenko,
and N.~Kwak}
\address{
University of Kansas, Lawrence, Kansas 66045}
\author{
S.~Anderson, Y.~Kubota, M.~Lattery, J.~J.~O'Neill, S.~Patton,
R.~Poling, T.~Riehle, V.~Savinov, and A.~Smith}
\address{
University of Minnesota, Minneapolis, Minnesota 55455}
\author{
M.~S.~Alam, S.~B.~Athar, Z.~Ling, A.~H.~Mahmood, H.~Severini,
S.~Timm, and F.~Wappler}
\address{
State University of New York at Albany, Albany, New York 12222}
\author{
A.~Anastassov, S.~Blinov,%
\thanks{Permanent address: BINP, RU-630090 Novosibirsk, Russia.}
J.~E.~Duboscq, K.~D.~Fisher, D.~Fujino,%
\thanks{Permanent address: Lawrence Livermore National Laboratory, Livermore, CA 94551.}
R.~Fulton, K.~K.~Gan, T.~Hart, K.~Honscheid, H.~Kagan, R.~Kass,
J.~Lee, M.~B.~Spencer, M.~Sung, A.~Undrus,%
$^{\addtocounter{footnote}{-1}\thefootnote\addtocounter{footnote}{1}}$
R.~Wanke, A.~Wolf, and M.~M.~Zoeller}
\address{
Ohio State University, Columbus, Ohio 43210}
 
\author{(CLEO Collaboration)}

\maketitle

\begin{abstract}
Using data taken with the CLEO II detector at the Cornell Electron Storage
Ring, we have determined the ratio of branching fractions:
$R_{\gamma} \equiv
\Gamma(\Upsilon({\rm 1S}) 
\rightarrow \gamma gg)/\Gamma(\Upsilon({\rm 1S}) \rightarrow ggg)
= (2.75 \pm 0.04(stat.) \pm 0.15(syst.))\%$.
From this ratio, we have determined the
QCD scale parameter $\Lambda_{\overline{MS}}$ (defined in the
modified minimal subtraction scheme) to be
$\Lambda_{\overline{MS}}= 233 \pm 11 \pm 59$ MeV, from which we
determine a value for the strong coupling constant
$\alpha_{\rm s}(M_{\Upsilon({\rm 1S})}) = 0.163 \pm 0.002 \pm 0.014$,
or $\alpha_{\rm s}(M_{Z}) = 0.110 \pm 0.001 \pm 0.007$.
\end{abstract}
\pacs{PACS numbers: 13.40.Dk, 13.25.+m, 14.40.Jz}

\section{Introduction}

The three primary decay modes of the $\Upsilon({\rm 1S})$ are three
gluons ($ggg$), a virtual photon ($\gamma^*$), or two gluons plus a photon
($\gamma gg$). We expect these decay widths to vary as:
$\Gamma_{ggg} \propto \alpha_{\rm s}^3$,
$\Gamma_{\gamma^*}\propto \alpha_{\rm em}^2$, and
$\Gamma_{\gamma gg} \propto  \alpha_{\rm em} \alpha_{\rm s}^2$,
respectively. From the ratio of decay rates:
\begin{equation}
R_{\gamma} \equiv \frac{\Gamma_{\gamma gg}}{\Gamma_{ggg}} \propto
\frac{N_{\Upsilon({\rm 1S})\to\gamma gg}}{N_{\Upsilon({\rm 1S})\to ggg}}\propto
                  \frac{\alpha_{\rm em}}{\alpha_{\rm s}},
\end{equation}
\noindent
one can determine a value for the strong coupling constant, $\alpha_{\rm s}$.
In this analysis we determined this ratio by measuring the number of direct
photon and three gluon events from a sample of $\Upsilon({\rm 1S})$ data.

\section{Detector, Data Sample, and Event Selection}
The CLEO~II detector is a general purpose solenoidal magnet spectrometer and
calorimeter. Elements of the detector, as well as performance characteristics,
are described in detail elsewhere \cite{r:CLEO-II}. For photons in the central
``barrel'' region of the CsI electromagnetic calorimeter the energy resolution
is given by
\begin{equation}
\frac{ \sigma_{\rm E}}{E}(\%) = \frac{0.35}{E^{0.75}} + 1.9 - 0.1E,
                                \label{eq:resolution1}
\end{equation}
where $E$ is the shower energy in GeV. The tracking system, time of flight
counters, and calorimeter are all contained within a 1.5 Tesla superconducting
coil.

The data used in this analysis were collected on the $\Upsilon({\rm 1S})$
resonance at a center-of-mass energy $E_{cm} = 9.46$ GeV, and from the
continuum region at a center-of-mass energy $E_{cm} = 10.52$ GeV, just below
the $\Upsilon({\rm 4S})$ resonance. The latter data set is used to subtract
out the nonresonant continuum events produced at $E_{cm} = 9.46$ GeV.
The event sample taken at the $\Upsilon({\rm 1S})$ energy corresponds to an
integrated luminosity of 62.5 pb$^{-1}$ acquired during two different running
periods.  The sample of continuum events chosen for our background studies
corresponds to an integrated luminosity of 91.3 pb$^{-1}$.

To obtain a clean sample of hadronic events, we selected those events
that had a minimum of three good charged tracks (to suppress contamination
from QED events), a total visible energy greater than 15\% of the total
center-of-mass energy (to reduce contamination from two-photon events
and beam-gas interactions), and an event vertex position 
consistent with the nominal $e^+e^-$ collision point to within
$\pm 5$ cm along the $e^+e^-$ axis ($z$) and $\pm$2 cm in the transverse
($r-\phi$) plane.
Backgrounds due to
radiative Bhabha events with a converted photon ($e^+e^-\to e^+e^-\gamma,
\gamma\to e^+e^-$) are reduced by requiring the total shower energy to
be at least 15\% of the available center-of-mass energy, but not more than
90\% of the available center-of-mass energy, as well as by rejecting events
that have thrust values approaching 1.0.

Applying these cuts, we obtained $1.43 \times 10^6$ events from the two
$\Upsilon({\rm 1S})$ data samples, collectively.

\section{The Inclusive Photon Spectrum}

\begin{figure}
\centering{\hbox{\psfig
{file=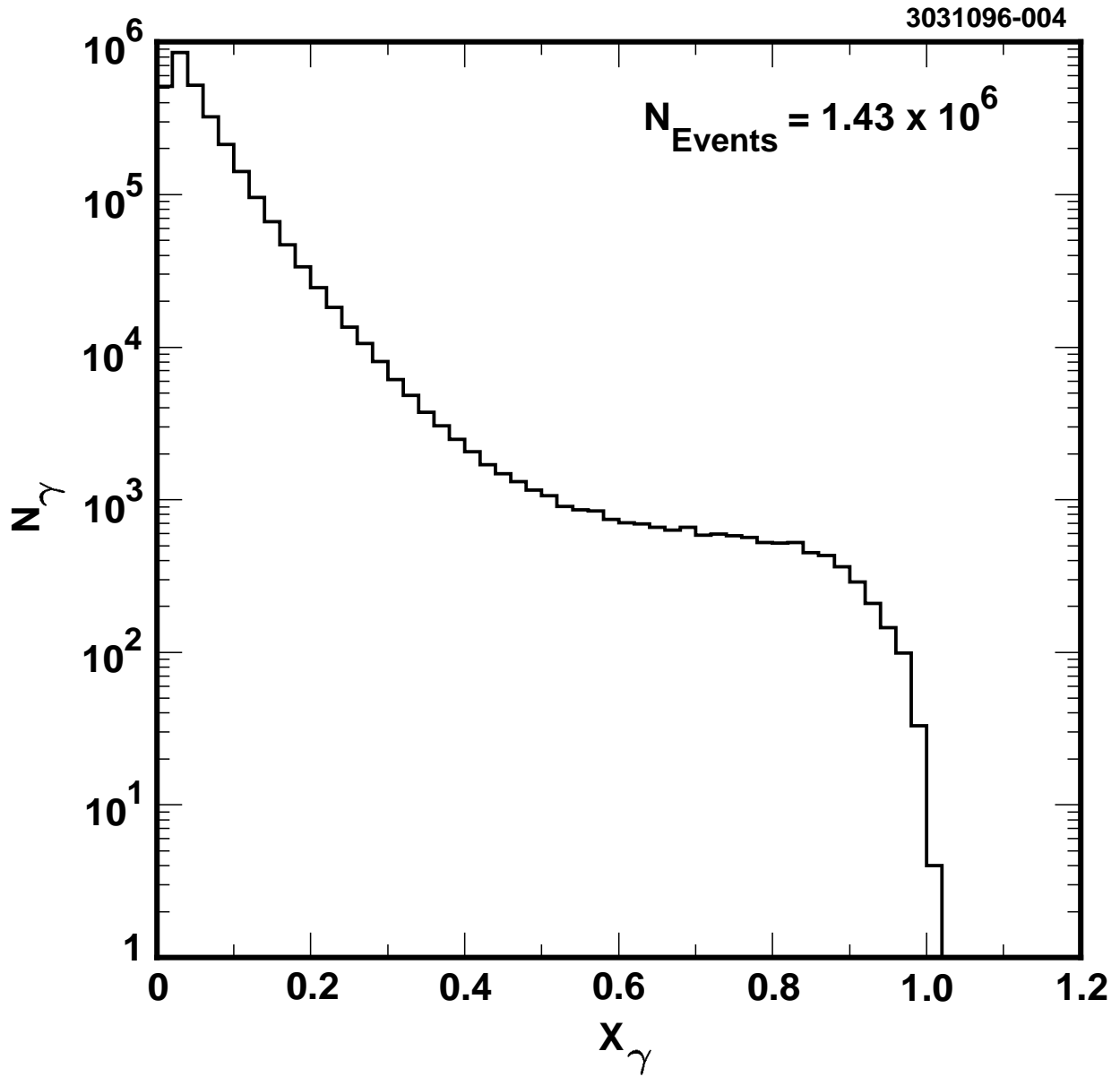,width=6.5truein}}}
\vskip 0.25truein
\caption[]
{The inclusive distribution of candidate photons as a function of scaled
 momentum $X_{\gamma} \equiv p_{\gamma}/E_{beam}$, from data taken at the
 $\Upsilon{\rm (1S)}$ center-of-mass energy.
 Note the log scale of the y-axis.}
\label{fig:1s-incl} 
\end{figure}

To obtain $R_{\gamma}$, we first compiled an inclusive photon spectrum from
the clusters of energy in the electromagnetic calorimeter.  Only photons from
the barrel region ($|\cos\theta_{\gamma}| < 0.7$, where $\theta_{\gamma}$ is
the polar angle of the shower) were considered.  Photon candidates were
required to be well separated from charged tracks and other photon candidates.
The lateral shower shape was required to be consistent with that expected from
a true photon. If the invariant mass of any two photon candidates fell within
15 MeV of the
$\pi^0$ mass, then both photons were rejected.  Photons produced in the decay
of a highly energetic $\pi^0$ would sometimes produce overlapping showers in
the calorimeter, creating a so-called merged $\pi^0$.  To remove this
background, an effective invariant mass was determined from the energy
distribution within a single electromagnetic shower. Showers whose effective
invariant masses were consistent with those from merged $\pi^0$'s were also
rejected. Figure~\ref{fig:1s-incl} shows the inclusive spectrum that results
from these cuts as a function of the scaled momentum variable,
$X_{\gamma} \equiv p_{\gamma}/E_{\rm beam}$.

\section{Background Sources}
\label{sect:bckgd} 
The dominant source of background photons is asymmetric $\pi^0$ decay.  To
remove this background, we developed a Monte Carlo simulation in which polar
angle and event selection effects were implicitly included.
Modulo isospin breaking effects, one expects similar kinematic distributions
between charged pions, which produce most of the charged tracks in
$\Upsilon({\rm 1S})$ hadronic decays,~\cite{r:Brown} and neutral pions,
which produce most of the background photons in the inclusive spectrum.
By measuring the ratio of the true $\pi^0$ momentum spectrum $dN/dX_{\pi^0}$ to
the charged track spectrum $dN/dX_{\pi^{\pm}}$, the charged tracks themselves
could then be used as a basis for simulating photons from $\pi^0$ decays.

We therefore estimated the background due to photons produced in neutral
meson decays as follows: for events that passed our selection criteria, we
measured the ratio of efficiency-corrected $\pi^0$'s to observed charged
tracks as a function of momentum. Then, assuming that the angular
distribution of $\pi^0$'s is the same as that for charged tracks,
the 3-momenta of the charged tracks were used to generate the expected
background spectrum from $\pi^0$ decays (with the correct angular correlations
implicit). The measured ratio provided the appropriate normalization.

\begin{figure}
 \centering{\hbox{\psfig
{file=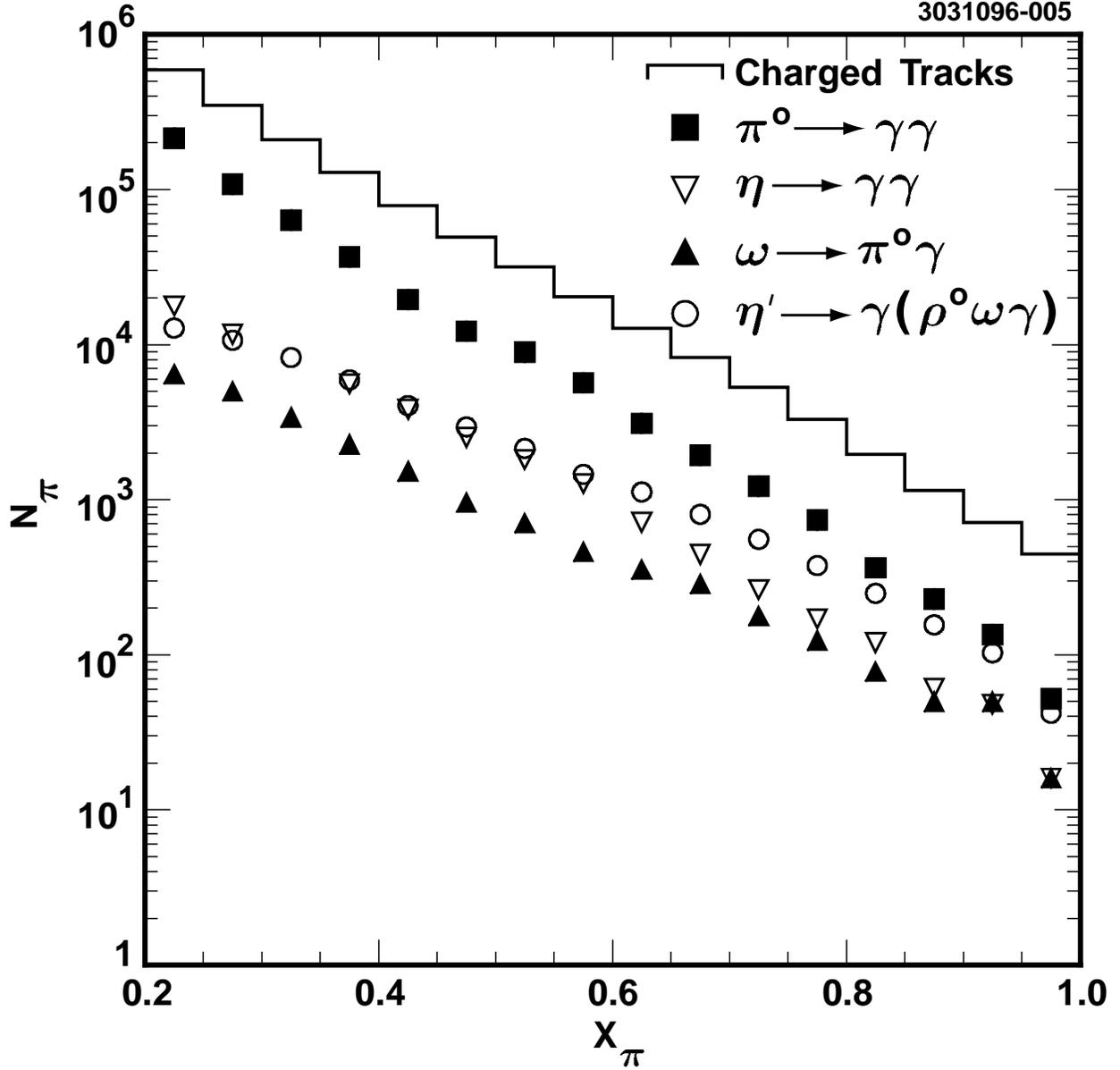,width=6.5truein}}}
\vskip 0.25truein
\caption[]
{Efficiency corrected 
$\pi^0$ momentum spectrum, Monte Carlo generated $\eta$, $\eta'$,
and $\omega$ spectra, and observed charged tracks' momentum spectrum as a
function of scaled particle momentum, $X_{\pi} = p_{\pi}/E_{\rm beam}$. In
this notation, `$\pi$' refers to any of the neutral spectra or charged tracks.}
\label{fig:charged:neutrals} 
\end{figure}

This approach had the advantage of being less model dependent than a Monte
Carlo event generator, as the ``generator'' in this method was the data itself.
It had the additional virtue that the absolute normalization of the
$\pi^0$ background was simply determined by the number of accepted events.
In addition to simulating the $\pi^0 \rightarrow \gamma \gamma$ background,
this technique was also used to account for
$\eta \rightarrow \gamma \gamma$, $\omega \rightarrow \pi^0 \gamma$,
and $\eta' \rightarrow \gamma [\rho^0, \omega, \gamma]$ contributions.
Figure~\ref{fig:charged:neutrals} illustrates the corrected momentum spectra
of these neutral mesons and charged tracks used to emulate their decays.

Contributions from long lived neutral hadrons (neutrons, anti-neutrons, and
${\rm K}^0_L$'s) can also produce showers in the calorimeter. We used the
LUND/JETSET 7.3~\cite{r:LUND} Monte Carlo simulation of $\Upsilon({\rm 1S})$
decays to estimate the number of long lived neutral hadrons in our event
sample and a detector simulation based on the GEANT~\cite{r:GEANT} package to
determine how often these ``residual showers'' would pass the photon selection
criteria.  It was found that these hadrons represented a small contribution,
not exceeding 3\% for any value of $X_{\gamma}$.

\begin{figure}
\centering{\hbox{\psfig
{file=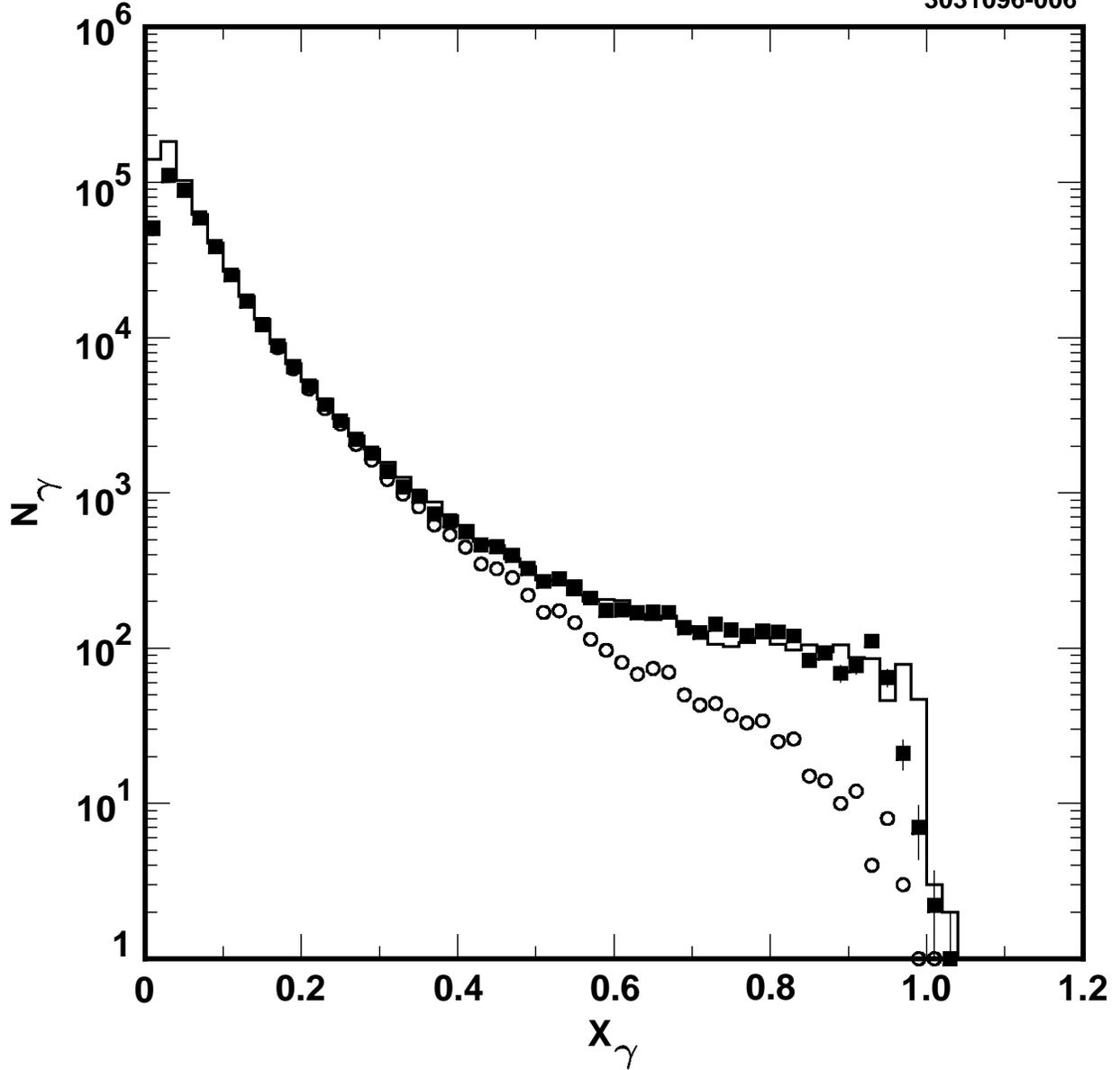,width=6.5truein}}}
\vskip 0.25truein
\caption[]
{A comparison of the inclusive photon spectrum from continuum data taken below
 the $\Upsilon(4{\rm S})$ resonance (histogram) with the simulated background
 spectrum from $\pi^0$'s, $\eta$'s, $\eta'$'s, and $\omega$'s produced by
 non-resonant processes at this energy. To illustrate the magnitude of the
 initial state radiative (ISR) correction, the simulated spectrum both with
 (dark squares) and without (open circles) the (Monte Carlo determined)
 radiative contribution are overlaid.}
\label{fig:dkpvcon-mom} 
\end{figure}

A test of this background simulation method was performed using data
collected from the continuum region, $E_{cm} = 10.52$ GeV. Using a set of
ratios for charged tracks to $\pi^0$'s, $\eta$'s, $\eta'$'s, and $\omega$'s
measured at this energy, we generated a photon spectrum and
compared it to the inclusive spectrum from the continuum. With the exception
of initial state radiation (whose contribution could be estimated from 
LUND and GEANT Monte Carlo simulations), the inclusive photon spectrum
and simulated photon spectrum should agree. Figure~\ref{fig:dkpvcon-mom} shows
this comparison.  We observe good agreement over the full range of
$X_{\gamma}$.\footnote{Note: the Monte Carlo simulation of initial state
radiation was not used as part of the final background subtraction. It is
included in Figure~\ref{fig:dkpvcon-mom} only to demonstrate that the
background contribution to the inclusive photon spectrum is well modeled.
Initial state radiation photons were automatically removed when we performed
a scaled continuum subtraction to remove nonresonant contributions to the
inclusive spectrum taken at $E_{cm} = 9.46$ GeV.}

\subsection{Subtractions and efficiencies}
\label{sect:subtract} 

This analysis has three major sources of background photons: (1) neutral
hadrons (specifically, $\pi^0$'s, $\eta$'s, $\eta'$'s, and $\omega$'s)
produced in $\Upsilon({\rm 1S})$ decay, (2) neutral hadrons produced in
nonresonant $e^+e^- \rightarrow q \bar{q} \gamma X$ processes, and (3)
radiative photons from the process $e^+e^- \rightarrow q \bar{q} \gamma$.  By
subtracting the $dN/dX_\gamma$ spectrum from the continuum data, scaled to
correct for the differences in luminosity and cross section, we remove
background from the latter two classes.  

The photon spectrum that we generated using charged particles collected at the
$\Upsilon({\rm 1S})$ energy simulates the spectrum from the first two
background classes combined, while the spectrum generated using charged
particles from the continuum sample simulates only the second class.  By
subtracting these two spectra after appropriate scaling, we isolate the
background spectrum of indirect photons from $\Upsilon({\rm 1S})$ decay.
Hence subtracting the resulting spectrum from the data removes the first class
of background.

\begin{figure}
 \centering{\hbox{\psfig
{file=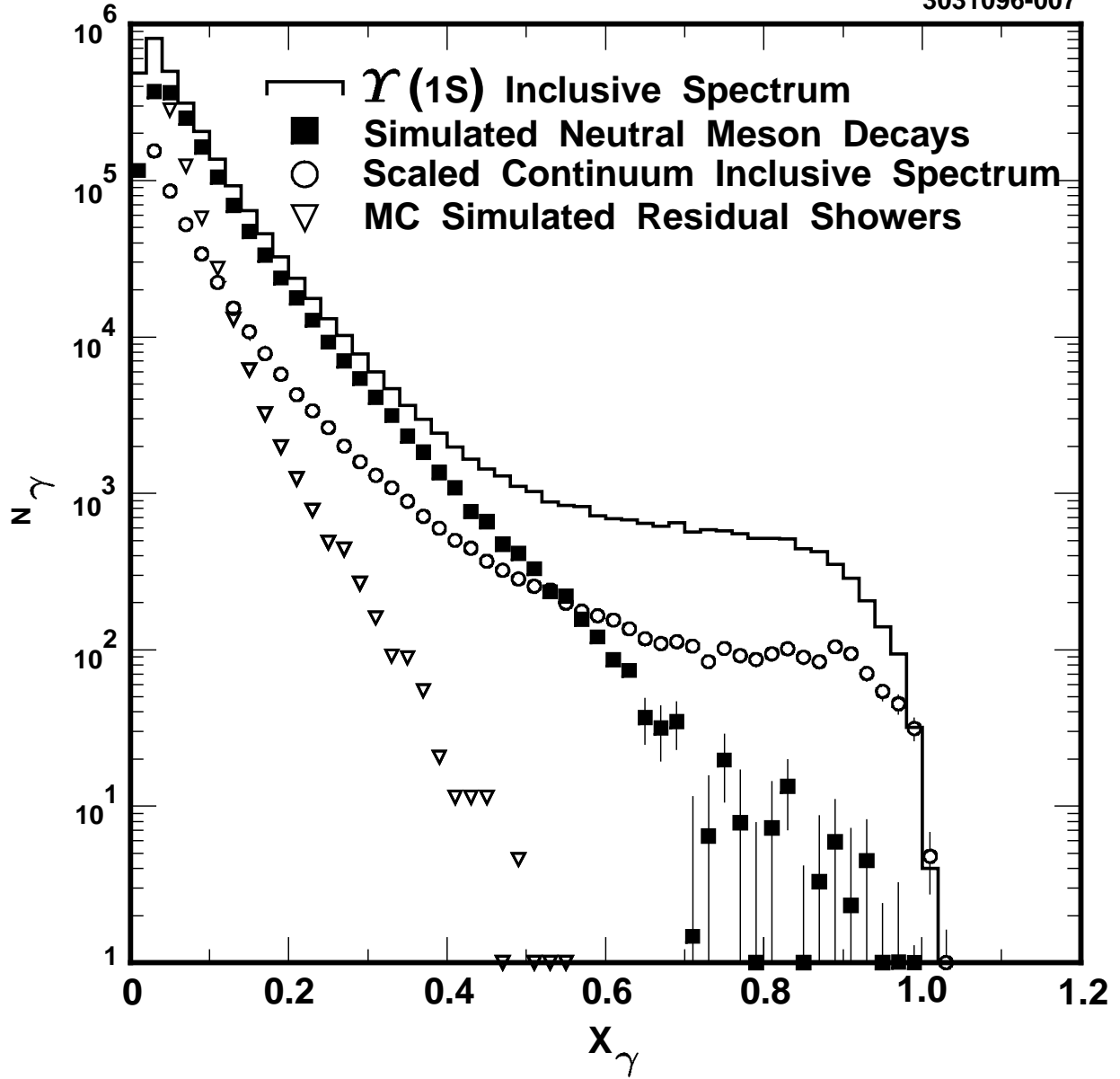,width=6.5truein}}}
\vskip 0.25truein
\caption[]
{The inclusive $X_\gamma$ spectrum (histogram) for data taken at the
 $\Upsilon({\rm 1S})$ resonance, along with background contributions due to
 non-resonant processes (open circles), resonant hadronic decays
 (dark squares), and other residual sources (inverted triangles).}
\label{fig:1sinc+bkgd} 
\end{figure}

Figure~\ref{fig:1sinc+bkgd} shows the inclusive $X_\gamma$ spectrum for
data taken on the $\Upsilon({\rm 1S})$ resonance,
with the different background contributions (non-resonant
hadronic~{\footnotesize\&}~radiative photons, resonant hadronic photons, and
residual showers) overlaid.  After subtracting these sources, what remained
of the inclusive $\Upsilon({\rm 1S})$ spectrum was identified as the direct
photon spectrum, $\Upsilon({\rm 1S}) \rightarrow \gamma gg$.

\begin{figure}
 \centering{\hbox{\psfig
{file=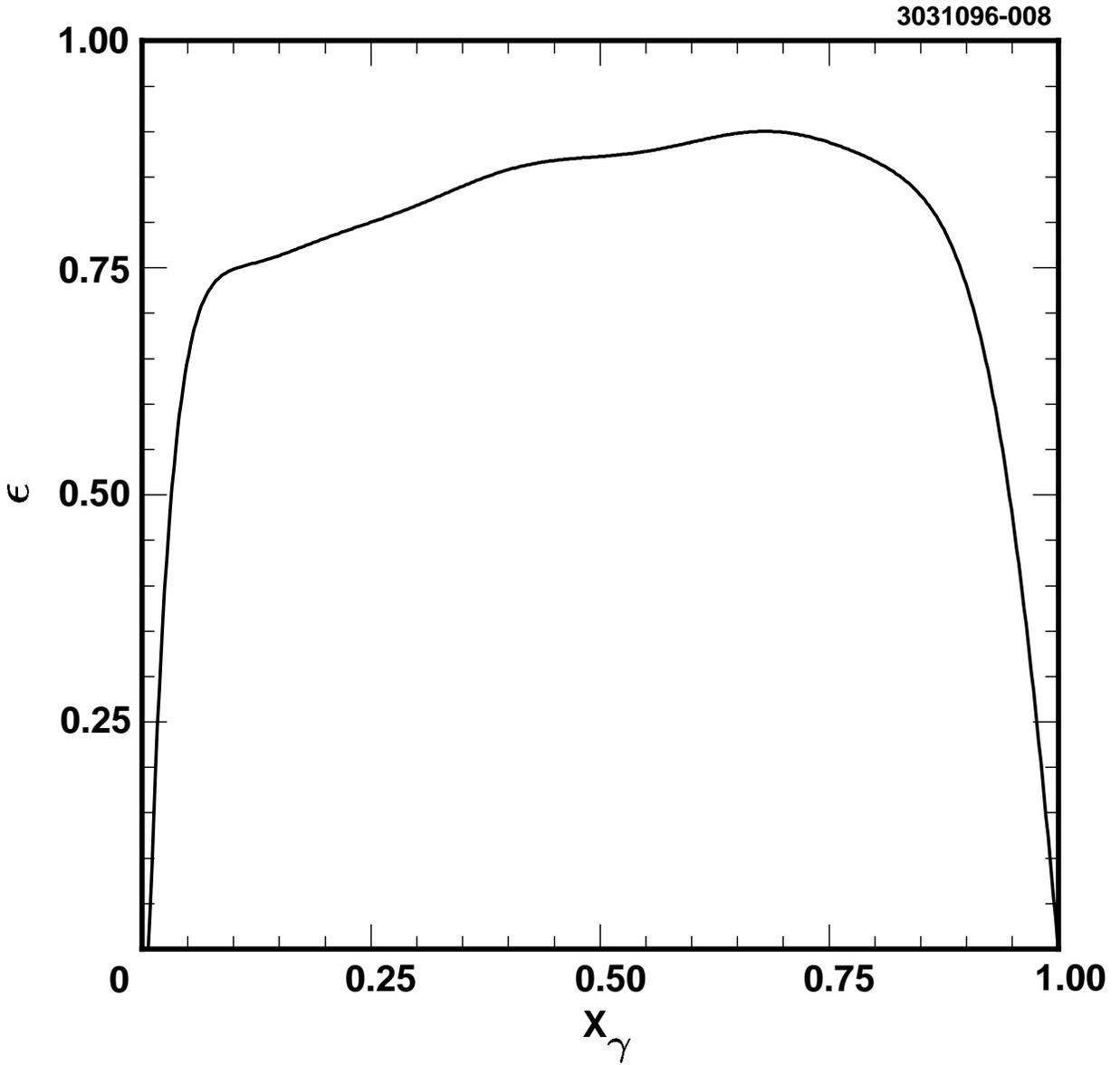,width=6.5truein}}}
\vskip 0.25truein
\caption[]
{The direct photon detection efficiency as a function of scaled photon
 momentum $X_{\gamma}$, determined by averaging the two Monte Carlo simulation
 models (see text).}
\label{fig:gggamm-eff}
\end{figure}

To compare our data with predictions for the shape of the direct photon
spectrum, we modified the theoretical distributions to account for
attenuation and distortion from the finite detection efficiency and
energy resolution. The most significant loss of direct photon events occurs at
the high-$X_{\gamma}$ region, arising from our requirement that an event have
at least three good charged tracks. Unfortunately, hadronization in this
kinematic regime is poorly understood. We considered two different $\Upsilon
\rightarrow \gamma gg$ Monte Carlo models with two different hadronization
schemes and used a photon detection efficiency from the average of the two
models (see Figure~\ref{fig:gggamm-eff}). The difference in efficiency between
the two models was incorporated into our systematic error.

Trigger efficiencies have been evaluated directly from the data by determining 
the fraction of events passing a minimum-bias trigger. This 
efficiency, for all values of photon momentum considered in this analysis,
exceeds 99\%.

\section{Comparison with Theoretical Models}
\subsection{Field Model}
As Figure~\ref{fig:1sinc+bkgd} illustrates, the inclusive $dN/dX_{\gamma}$
distribution increases rapidly in the low $X_{\gamma}$ region.  This is
due primarily to an overwhelming number of photons produced in 
$\pi^0 \to \gamma\gamma$ decays. However, to extract the total number
of $\Upsilon({\rm 1S}) \rightarrow gg\gamma$ events and obtain $R_{\gamma}$,
we needed to integrate this spectrum along the entire scaled momentum axis. It
was therefore necessary to rely on a model which fit well to that portion of
the photon spectrum where the signal photons were clearly observable so that
an extrapolation into the lower momentum, higher background region could be
performed confidently.  A number of attempts have been made to predict the
shape of this spectrum~\cite{r:lowest-qcd,r:Photiadis,r:Catani,r:Field}.
In this analysis, we employed the model by Field~\cite{r:Field} for our
integration purposes.

\begin{figure}
\centering{\hbox{\psfig
{file=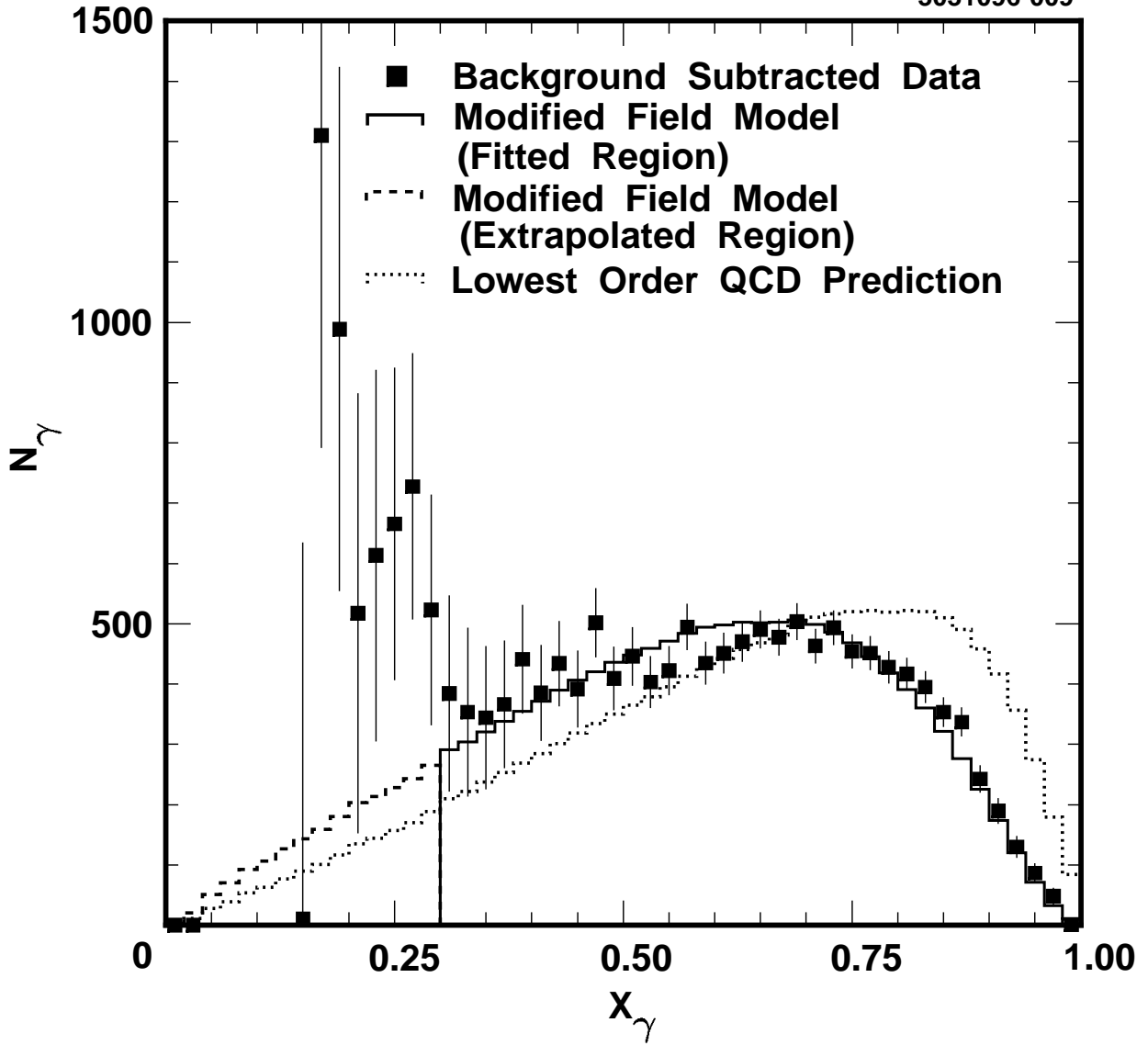,width=6.5truein}}}
\vskip 0.25truein
\caption[]
{The background subtracted (i.e., direct) photon spectrum
(dark squares). The data points in the region $0.3 < X_{\gamma} < 0.98$ are
fit to Field's model (histogram).  The only free parameter in the fit was the
overall normalization. The fit to Field's model, extrapolated into the low
momentum region, is also shown (dashed line) as well as the lowest order QCD
prediction (dotted line) over the full kinematic regime, which has been
normalized to the same area as the Field model. The errors shown are purely
statistical. Data points in the region
$X_{\gamma} < 0.3$ appear systematically shifted above the Field line,
however, a one sigma shift in the magnitude of the simulated background line
in Figure~\ref{fig:1sinc+bkgd} would drastically alter the distribution at the
low $X_{\gamma}$ end. Systematic errors, had they been included on these
points, would be well off scale.}
\label{fig:Field} 
\end{figure}

Figure~\ref{fig:Field} shows our photon spectrum with the background sources
subtracted.  To determine the number of direct photon events, $N_{\gamma gg}$,
from this spectrum, the data points in the region $0.30 < X_{\gamma} < 0.98$
were fit to the modified (i.e., efficiency-attenuated and energy smeared)
Field model; the only free parameter in this fit was the overall normalization.
For comparison purposes, the modified lowest order QCD prediction,\footnote{In
the lowest order QCD prediction, the $\Upsilon$ system is treated as
ortho-positronium decaying into three photons.} normalized to the same area as
the Field model, has been overlaid.
According to Field's model, about 85\% of the direct photons that are produced
in $\gamma gg$ decays lie within this portion of the momentum spectrum.
According to our averaged detection efficiency of Figure~\ref{fig:gggamm-eff},
about 15\% of those events are rejected by our shower and event-selection cuts.

To determine the fraction of direct photons within our fiducial acceptance,
we used a Monte Carlo simulation of the direct photon events, incorporating
the QCD predictions of Koller and Walsh\cite{r:kol-walsh} 
for the photon 
angular distributions as a function of momentum.  
According to their model, roughly
67\% of the direct photons fall within our fiducial acceptance, 
$|cos\theta|<0.7$.  
Thus, our subtracted spectrum, within the limits of the fit and
our fiducial acceptance,
represents approximately 48\% of the total direct photons produced in the
$\Upsilon({\rm 1S})$ data sample.

After integration of the fitted Field distribution in Figure~\ref{fig:Field}
and corrections for finite acceptance, our data yields a total number of
$\Upsilon({\rm 1S}) \rightarrow \gamma gg$ decays, 
$N_{\gamma gg} = (2.652 \pm 0.038) \times 10^4$.

To determine the number of three gluon events $N_{ggg}$ from the number of
observed $\Upsilon({\rm 1S})$ hadronic events $N^{\Upsilon({\rm 1S})}_{had}$,
we first determined the number of continuum events under the 
$\Upsilon({\rm 1S})$ resonance
($N^{\Upsilon({\rm 1S})}_{cont}$) from the observed number of 
$\Upsilon({\rm 4S})$ continuum events $N^{\Upsilon({\rm 4S})}_{cont}$, 
accounting for the dependence of the cross-section on $E_{cm}^2(\equiv S)$:
\begin{equation}
N^{\Upsilon({\rm 1S})}_{cont} = N^{\Upsilon({\rm 4S})}_{cont} \cdot
                 \frac{ {\cal L}_{\Upsilon({\rm 1S})} }{ {\cal L}_{\rm cont} }
                 \cdot \frac{ S_{\rm cont} }{ S_{\Upsilon({\rm 1S})} }
                 = 3.21 \times 10^{5},
\end{equation}
\noindent
Next, we estimated the number of vacuum-polarization events $N_{vp}$ using
the $\Upsilon({\rm 1S}) \rightarrow \mu^+ \mu^-$ branching fraction
$B_{\mu \mu} = 0.0248$~\cite{r:PRD}, and
$R_{\Upsilon({\rm 1S})} = \sigma(e^+e^- \rightarrow \Upsilon({\rm 1S})
\rightarrow q{\bar q}) / \sigma(e^+e^- \rightarrow \Upsilon({\rm 1S})
\rightarrow \mu^+ \mu^-) = 3.46 \pm 0.14$~\cite{r:Albrecht-92}:
\begin{equation}
N_{vp} = R_{\Upsilon({\rm 1S})} \cdot B_{\mu \mu} \cdot
         \frac{ N^{\Upsilon({\rm 1S})}_{had} }{ (1 - 3 B_{\mu \mu}) }
         = 1.27 \times 10^{5}
\end{equation}
\noindent
From these values and Monte Carlo determined efficiencies for the various
event types to pass our event selection selection criteria (see
Table~\ref{tab:evnt-eff}), we determined

\begin{equation}
N_{ggg} = (N^{\Upsilon({\rm 1S})}_{had} - N^{\Upsilon({\rm 1S})}_{cont}
           - N_{vp}\cdot(\epsilon_{vp}/\epsilon_{had}) 
           - N_{\gamma gg}\cdot(\epsilon_{\gamma gg}) )/\epsilon_{ggg} =
           (9.657 \pm 0.010) \times 10^5.
\end{equation}
From these values we obtained a value for
$R_{\gamma}$:
\begin{equation}
R_{\gamma} = \frac{ N_{\gamma gg} }{ N_{ggg} } = 2.75 \pm 0.04\%
\end{equation}

\subsection{Catani and Hautmann Modification to $\gamma gg$ Spectra}
\begin{figure}
\centering{\hbox{\psfig
{file=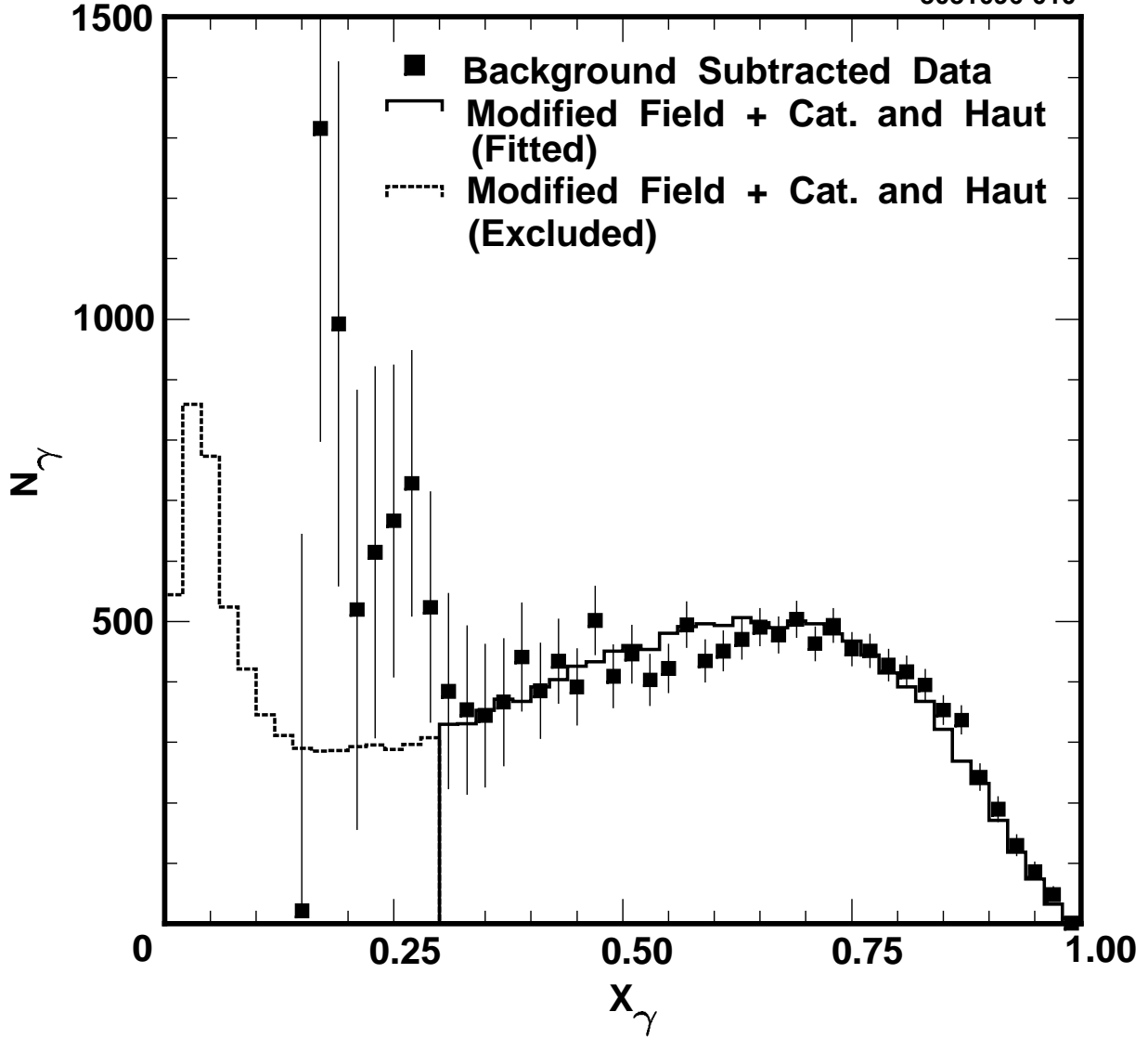,width=6.5truein}}}
\vskip 0.25truein
\caption[]
{The background subtracted photon spectrum (dark squares) fit to the Field
 distribution with the added fragmentation component predicted by Catani and
 Hautmann. Again, the errors shown on the data points are purely statistical.}
\label{fig:Catani} 
\end{figure}

Catani and Hautmann~\cite{r:Catani} assert that in order to determine the
total photon spectrum from $\Upsilon$(1S) decays
one must also consider fragmentation photons emitted from final--state
light quarks produced in the initial heavy quarkonia decay.
To properly measure $\alpha_{\rm s}$,
they claim, one must account for these additional photons, both in the shape
of the spectrum, as well
as in the QCD equations from which $\alpha_{\rm s}$ is
extracted.  
They also
provide a leading order estimate of the shape of the
prompt photon spectrum due to this fragmentation component.  In our analysis,
we added this same component to the direct spectrum predicted by Field,
modified the resulting spectrum for efficiency and energy
resolution, and fit this distribution to our data using essentially the
same method to determine $N_{\gamma gg}$ as described above.
From that distribution, we measured $R_{\gamma} = (2.72 \pm 0.04)\%$
(see Figure~\ref{fig:Catani}). Clearly, we do not yet have the requisite
experimental sensitivity needed to verify the Catani and Hautmann model.

\section{Systematic Errors}
Table~\ref{tab:syserrors} summarizes the systematic errors studied in this
analysis and their estimated effect upon $R_{\gamma}$.\footnote{Perhaps the 
largest potential errors arises from modeling the momentum and angular
distributions of the initial partons (i.e., the Field and Koller-Walsh
predictions). Although these are not explicitly folded into our overall
systematic error, it should be pointed out that our results are sensitive to
these predictions.} The tracking efficiency and multiplicity modeling
uncertainty was obtained by applying the two Monte Carlo models (with their
different hadronization schemes) separately, as opposed to their average. 
Including the $\pi^0$ veto reduces our statistical errors in the low
$X_{\gamma}$ region, but also adds to the uncertainty in our ability to
accurately simulate this cut.  The difference
between the value of $R_{\gamma}$ obtained by applying the $\pi^0$ veto and
the value obtained when we did not apply this veto constituted our second
largest systematic uncertainty in $R_{\gamma}$. By scaling the secondary
photon spectrum by $\pm 5\%$, we obtained the systematic error due to our
uncertainty in the overall normalization of the secondary photon spectrum.
We also compared results by using a different subtraction technique
in which the non-resonant radiative contribution was subtracted using
Monte Carlo simulated continuum events generated at the $\Upsilon({\rm 1S})$
center-of-mass energy. This allowed us to extract a value of $R_{\gamma}$
independent of any non-resonant data taken at energies other than 9.46 GeV.
The estimated uncertainty in the number of three gluon events
can be directly translated to an uncertainty in $R_{\gamma}$.
To check against possible
systematic effects due to different running conditions, we analyzed the two
$\Upsilon({\rm 1S)}$ data samples separately.
Finally, we included the total error (statistical and systematic, combined in
quadrature) quoted by ARGUS in their measurement of the ratio
of hadronic to muonic cross-sections, $R_{\Upsilon({\rm 1S})}$.

Table~\ref{tab:comps} compares the results of this analysis with those
obtained by previous experiments in which the observed number of
$\Upsilon({\rm 1S}) \rightarrow \gamma gg $ events were also determined using
Field's model.

\section{Extraction of QCD parameters}
We now relate the value of $R_\gamma$ to the fundamental QCD parameters which
we wish to measure, following
Sanghera~\cite{r:Sanghera}.

The decay width $\Upsilon \rightarrow \gamma gg$ has been calculated by Lepage
and Mackenzie~\cite{r:Mack-Lep} in terms of the coupling strength at the
energy scale characterizing this decay process, $\alpha_{\rm s}(M_{\Upsilon}$):
\begin{equation}
\frac{\Gamma(\Upsilon \rightarrow \gamma gg)}
     {\Gamma(\Upsilon \rightarrow \mu^+\mu^-)}
     = \frac{8(\pi^2 - 9)}{9 \pi \alpha_{QED}} \alpha_{\rm s}^2(M_{\Upsilon})
       \Biggl[1 + (3.7 \pm 0.4)\frac{\alpha_{\rm s}(M_{\Upsilon})}{\pi}
       \Biggr]. \label{eq:qcd0}
\end{equation}
\noindent
Expressing this ratio in terms of a leading-order power series in
$\alpha_s(\mu)$, we have:
\begin{equation}
 \frac{\Gamma(\Upsilon \rightarrow \gamma gg)}
      {\Gamma(\Upsilon \rightarrow \mu^+\mu^-)}
      = A_{\gamma}\biggl({\alpha_{\rm s}(\mu)\over \pi}\biggr)^2 +
        A_{\gamma}\biggl({\alpha_{\rm s}(\mu)\over \pi}\biggr)^3
                  \biggl[2 \pi b_0
                         ln\biggl({\mu^2 \over M_{\Upsilon}^2}\biggr) +
                         (3.7 \pm 0.4)
                   \biggr], \label{eq:qcd1}
\end{equation}
\noindent
where $ A_{\gamma} = {8\pi (\pi^2-9) \over 9\alpha_{QED}}$,
$b_0=(33-2n_f)/12\pi$, and $n_f$ is the number of light quark flavors
which participate in the process ($n_f=4$ for $\Upsilon({\rm 1S})$ decays).

Similarly, the decay width $\Upsilon \rightarrow ggg$ has been calculated
by Bardeen et al.~\cite{r:Bardeen} and expressed by Lepage
et al.~\cite{r:Brod-Lep-Mack,r:Mack-Lep-PRL} as:
\begin{equation}
 \frac{\Gamma(\Upsilon \rightarrow ggg)}
      {\Gamma(\Upsilon \rightarrow \mu^+\mu^-)}
      = \frac{10(\pi^2 - 9)}{81 \pi e_b^2}
        \frac{\alpha_{\rm s}^3(M_{\Upsilon})}{\alpha^2_{QED}}
        \Biggl[1 + \frac{\alpha_{\rm s}(M_{\Upsilon})}{\pi}
              [2.770(7)\beta_0 - 14.0(5)] + \cdot \cdot \cdot \Biggr],
        \label{eq:qcd101}
\end{equation}
where $\beta_0 = 11 - ({2 \over 3})n_f$, and $e_b = -{1 \over 3}$, the charge
of the b quark.  Again,
we can express
this in terms of the renormalization scale:
\begin{equation}
 \frac{\Gamma(\Upsilon \rightarrow ggg)}
      {\Gamma(\Upsilon \rightarrow \mu^+\mu^-)}
      = A_g\biggl({\alpha_{\rm s}(\mu)\over \pi}\biggr)^3 +
        A_g\biggl({\alpha_{\rm s}(\mu)\over \pi}\biggr)^4
           \biggl[3 \pi b_0 ln\biggl({\mu^2 \over M_{\Upsilon}^2}\biggr) -
           \biggl({2 \over 3}\biggr)B_f n_f + B_i \biggr], \label{eq:qcd2}
\end{equation}
\noindent
where $ A_g = {10\pi^2(\pi^2-9) \over 81 e_b^2} {1 \over \alpha_{QED}^2}$,
$ B_f = 2.770 \pm 0.007$, and $B_i = 16.47 \pm 0.58$.

The strong coupling constant $\alpha_{\rm s}$ can be written as a function of
the basic QCD parameter $\Lambda_{\overline{MS}}$, defined in the modified
minimal subtraction scheme~\cite{r:PRD},
\begin{equation}
    \alpha_{\rm s}(\mu) = {1 \over b_0 ln(\mu^2/\Lambda_{\overline{MS}}^2)}
                          \Biggl(1 - {b_1 \over b_0^2}
                          {ln(ln({\mu^2}/{\Lambda_{\overline{MS}}^2})) \over
            ln({\mu^2}/{\Lambda_{\overline{MS}}^2})}\Biggr) \label{eq:qcd3}
\end{equation}
\noindent
where $b_1 = {153 - 19n_f \over 24\pi^2}$.

Note that the scale dependent QCD equations~(\ref{eq:qcd1})
and~(\ref{eq:qcd2}) are of finite order in $\alpha_{\rm s}$. If these equations
were solved to all orders, then they could in principle be used to determine
$R_{\gamma}$ independent of the renormalization scale.  Because we are
dealing with calculations that are of finite order, the question of an
appropriate scale must be addressed.

The renormalization scale may be defined in terms of the center of mass
energy of the process, $ \mu^2= f_{\mu} E_{cm}^2$, where $f_{\mu}$ is some
positive value.  But QCD does not tell us {\it a priori} what $f_{\mu}$ should
be. One possibility would be to define $\mu=E_{cm}$; that is $f_{\mu}=1$. A
number of phenomenological
prescriptions~\cite{r:Sanghera,r:Brod-Lep-Mack,r:Grunberg,r:Stevenson} have
been proposed in an attempt to ``optimize'' the scale.  However, each of these
prescriptions yields scale values which, in general, vary greatly with the
experimental quantity being measured~\cite{r:Sanghera}. 

\begin{figure}
\centering
\centerline{\hbox{\psfig
 {file=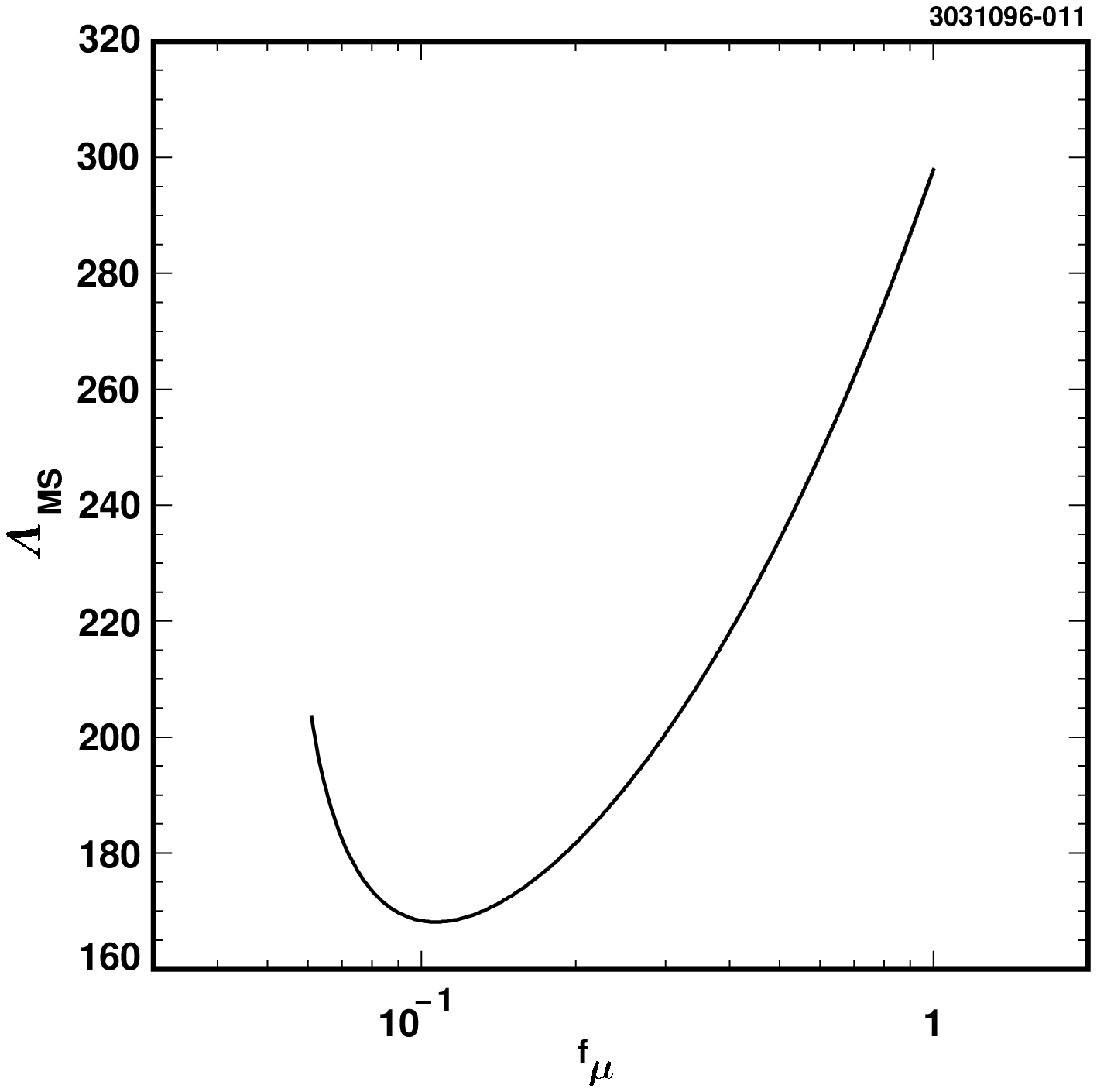,width=6.5truein}}}
\vskip 0.25truein
\caption[]
{$\Lambda_{\overline{MS}}$ as a function of scale parameter $f_{\mu}$,
 as governed by the functional dependence of $R_{\gamma}$ on
 $\Lambda_{\overline{MS}}$ and $f_\mu$.}
\label{fig:QCD} 
\end{figure}

In
this analysis, we have determined
$\Lambda_{\overline{MS}}$ over a range of scale values.
This was done by comparing our measured value of $R_{\gamma}$ with the ratio
of equations~(\ref{eq:qcd1}) and~(\ref{eq:qcd2}) in which $\alpha_{\rm s}$ was
replaced by the expression in equation~(\ref{eq:qcd3}), thereby providing a
relationship between $\Lambda_{\overline{MS}}$, $f_{\mu}$, and $R_{\gamma}$.
Thus for each assumed value of $f_{\mu}$, $\Lambda_{\overline{MS}}$ was
numerically determined as a function of $R_{\gamma}$.  The resulting
$\Lambda_{\overline{MS}}$ versus $f_{\mu}$ dependence is shown in
Figure~\ref{fig:QCD}. This dependence was parameterized by the form
\begin{equation}
      \Lambda_{\overline{MS}}(f) = \Lambda_{\overline{MS}}(f_0)
      + (c_1)ln\biggl({f\over f_0}\biggr)+
          (c_1+c_2)ln^2\biggl({f \over f_0}\biggr)
\end{equation}

\noindent
where $f_0$ is the value of $f_{\mu}$ around which $\Lambda_{\overline{MS}}$
is minimally dependent on $f_{\mu}$, given by
$(\partial \Lambda_{\overline{MS}} / \partial f_{\mu} = 0)$.  In this
analysis, we determined $f_0 = 0.107$, $\Lambda_{\overline{MS}}(f_0)$ = 
168.62 MeV, $c_1 = 7.74$, $c_2 = 14.68$. 

By parameterizing the results of the analysis in this manner, one can easily
extract QCD parameters at any scale within the range of the parameterization,
$0.10 \leq f_{\mu} \leq 1.0$, and compare with other results.  For example,
the mean value between $\Lambda_{\overline{MS}}(f_{\mu} = 0.107)$ (where
$\Lambda_{\overline{MS}}$ is a minimum), and
$\Lambda_{\overline{MS}}(f_{\mu} = 1.0)$, is $233 \pm 11 \pm 59$ MeV. The
uncertainty of the parameterization due to the theoretical uncertainties of
the parameters in equations~(\ref{eq:qcd1}) and~(\ref{eq:qcd2}) has been 
included in the systematic error of $\Lambda_{\overline{MS}}$.  Substituting
this value for $\Lambda_{\overline{MS}}$ into 
equation~(\ref{eq:qcd3}), and using
$\mu = M_{\Upsilon({\rm 1S})}$ we find for $\alpha_{\rm s}$:
\begin{equation}
\alpha_{\rm s}(M_{\Upsilon({\rm 1S})}) = 0.163 \pm 0.002 \pm 0.009 \pm 0.010,
\end{equation}
\noindent
where the additional error of $\pm 0.010$ arises from the difference,
about 64 MeV, 
between the mean value of $\Lambda_{\overline{MS}}$ and the values at each of
the parameterization limits, $f_{\mu}$ = 0.107 and $f_{\mu}$ = 1.0.
Extrapolating this result to $\mu = M_{Z}$, and assuming
continuity of $\alpha_{\rm s}$ across the 5 flavor continuum
threshold~\cite{r:Marciano} (which implies that $\Lambda_{\overline{MS}}$ is
a step function across the 5 flavor threshold), we obtain from
equation~(\ref{eq:qcd3}):
\begin{equation}
\alpha_{\rm s}(M_{Z}) = 0.110 \pm 0.001 \pm 0.004 \pm 0.005.
\end{equation}
\noindent
This result is lower, although in
acceptable 
agreement with the average value of $\alpha_{\rm s}(M_{Z}) = 0.122 \pm
0.007$ presently quoted by the Particle Data Group\cite{r:PRD}.
It is worth noting that the value
of $\Lambda_{\overline{MS}}$ obtained by previous
experiments~\cite{r:Albrecht-87,r:Bizeti} using the fixed-scale procedure of
ref.~\cite{r:Brod-Lep-Mack} is in agreement with our value at $f_0$ where the
scale dependence on $\Lambda_{\overline{MS}}$ becomes minimal.

\medskip
\centerline{\bf ACKNOWLEDGMENTS}
\medskip
We gratefully acknowledge the effort of the CESR staff in providing us with
excellent luminosity and running conditions.
J.P.A., J.R.P., and I.P.J.S. thank
the NYI program of the NSF,
M.S thanks the PFF program of the NSF,
G.E. thanks the Heisenberg Foundation,
K.K.G., M.S., H.N.N., T.S., and H.Y. thank the
OJI program of DOE,
J.R.P, K.H., and M.S. thank the A.P. Sloan Foundation,
and A.W., and R.W. thank the
Alexander von Humboldt Stiftung
for support.
This work was supported by the National Science Foundation, the
U.S. Department of Energy, and the Natural Sciences and Engineering Research
Council of Canada.

\clearpage  
\begin{table}[h]
\caption[]{\label{tab:evnt-eff} Monte Carlo Determined Event Efficiencies}
\begin{tabular}{ccc}
 Event Type              &   Symbol               &  Efficiency \\
\hline
 three gluon             & $\epsilon_{ggg}$       & 0.9938 \\
 `generic' $\Upsilon({\rm 1S})$ \ hadronic & $\epsilon_{had}$      & 0.9985 \\
 vacuum polarization     & $\epsilon_{vp}$        & 0.9480 \\
 direct photon           & $\epsilon_{\gamma gg}$ & 0.9419 \\
\end{tabular}
\end{table}

\begin{table}[h]
\caption[]{\label{tab:syserrors} Systematic Errors}
\begin{tabular}{lr}
Uncertainty Source                            &  $\delta R_{\gamma} (\%)$  \\
\hline
 Tracking efficiency and multiplicity modeling   &  $0.12$ \\
 $\pi^0$ veto            &  $0.07$                    \\
 continuum subtraction     &  $0.04$                    \\
 $\epsilon_{ggg}$   &  $0.03$                    \\
 pseudo--photon spectrum           &  $0.03$                    \\
 Luminosity and $E_{cm}$ scaling          &  $0.02$                    \\
 $\Upsilon({\rm 1S})$ data samples used separately &  $0.02$    \\
 $\delta R_{\Upsilon({\rm 1S})}$   & $0.01$                     \\
\end{tabular}
\end{table}

\begin{table}[h]
\caption[]{\label{tab:comps} Comparison with Other Experiments}
\begin{tabular}{cc}
 Experiment                 & $R_{\gamma} (\%)$        \\
\hline
 CLEO 1.5~\cite{r:Csorna}   &  $2.54 \pm 0.18 \pm0.14$ \\
 ARGUS~\cite{r:Albrecht-87} &  $3.00 \pm 0.13 \pm0.18$ \\
 Crystal Ball~\cite{r:Bizeti}       &  $2.7  \pm 0.2  \pm 0.4$ \\
 This measurement           &  $2.75 \pm 0.04 \pm0.15$ \\ 
\end{tabular}
\end{table}

\clearpage  

\begin{thebibliography}{99}

\bibitem{r:CLEO-II} CLEO Collab., Y. Kubota $et~al$, ``The CLEO-II detector'',
                    Nucl. Instr. Meth. {\bf A320}, 66 (1992).

\bibitem{r:lowest-qcd} S.J. Brodsky, T.A.DeGrand, R.R. Horgan and D.G. Coyne,
                       Phys. Lett. {\bf B73} (1978) 203; K. Koller, and T.
                       Walsh, Nucl. Phys. {\bf B140}, (1978) 449.

\bibitem{r:Photiadis} D.M. Photiadis, Phys. Lett. {\bf B164} (1985) 160.

\bibitem{r:Catani} S. Catani and F. Hautmann, Nucl. Phys.
                   {\bf B}, Proc. Suppl. {\bf 39BC}, 359 (1995).

\bibitem{r:Field} R.D. Field, Phys. Lett. {\bf B133} (1983) 248.

\bibitem{r:Brown} David N. Brown, Ph.D. Dissertation, Purdue University
                  (1992), unpublished.

\bibitem{r:LUND} S. J. Sjostrand, LUND 7.3, CERN-TH-6488-92 (1992).

\bibitem{r:GEANT} R. Brun {\it et. al.}, GEANT v. 3.14, CERN Report No.
                  CERN CC/EE/84-1 (1987).


\bibitem{r:kol-walsh} K. Koller, and T. Walsh, ref~\cite{r:lowest-qcd}.

\bibitem{r:PRD} Review of Particle Properties, Phys. Rev. {\bf D54}, (1996).

\bibitem{r:Albrecht-92} H. Albrecht et al., ARGUS Collab., Z. Phys. {\bf C54}
                        (1992) 13.

\bibitem{r:Csorna} S.E. Csorna et al., CLEO Collab., Phys. Rev. Lett. {\bf 56}
                   (1986) 1222.

\bibitem{r:Albrecht-87} H. Albrecht et al., ARGUS Collab., Phys. Lett.
                        {\bf B199} (1987) 291.

\bibitem{r:Bizeti} A. Bizzeti et al., CRYSTAL BALL Collab., Phys. Lett.
                   {\bf B267} (1991) 286.

\bibitem{r:Sanghera} S. Sanghera, Int'l Journal of Mod. Phys. {\bf A9},
                     (1994) 5743, and S. Sanghera, Ph. D. Thesis, Carleton
U. (1991), unpublished, and S. Sanghera, private communication.

\bibitem{r:Mack-Lep} P. B. Mackenzie and G. Peter Lepage, in {\em Perturbative
                     Quantum Chromodynamics}, Conf. Proceed., Tallahassee,
                     1981, AIP, New York, 1981.


\bibitem{r:Bardeen} W. A. Bardeen et al., Phys. Rev. {\bf D18} (1978) 3998.

\bibitem{r:Brod-Lep-Mack} S. J. Brodsky, G. P. Lepage and P. B. Mackenzie,
                          Phys. Rev. {\bf D28} (1983) 228.

\bibitem{r:Mack-Lep-PRL} P. B. Mackenzie and G. Peter Lepage, Phys. Rev.
                         Lett. {\bf 47} (1981) 1244.

\bibitem{r:Grunberg} G. Grunberg, Phys. Lett. {\bf B95} (1980) 70.

\bibitem{r:Stevenson} P.M. Stevenson, Phys. Rev.  {\bf D23} (1981) 2916.


\bibitem{r:Marciano} W. J. Marciano, Phys. Rev. {\bf D29} (1984) 580.

\bibitem{r:Bethke} S. Bethke, Proc. International Conf. on High Energy Phys.,
                   Dallas TX, August 1992.

\bibitem{r:Benvenuti} A. C. Benvenuti, BCDMS Collab., Phys. Lett {\bf B223}
                      (1989) 490.

\end{thebibliography}
\end{document}